\documentclass[apj,twocolumn]{openjournal}
\usepackage{amsmath}
\usepackage{booktabs}
\usepackage[dvipsnames]{xcolor} 
\usepackage[breaklinks,colorlinks,citecolor=blue,urlcolor=blue,linkcolor=blue]{hyperref}
\usepackage{orcidlink}
\usepackage{listings}
\usepackage{color}
\usepackage{soul}
\usepackage{amsmath}
\usepackage{xspace}
\usepackage{graphicx}
\usepackage{enumitem}
\usepackage{amssymb}
\usepackage{xifthen}
\usepackage{hyperref}
\usepackage[normalem]{ulem}
\usepackage{multirow}
\usepackage[hang,flushmargin]{footmisc} 
\usepackage{styles/todo_notes}
\usepackage{styles/abbreviations}
\usepackage[utf8]{inputenc} 
\usepackage[T1]{fontenc}    
\usepackage{hyperref}       
\usepackage{url}            
\usepackage{booktabs}       
\usepackage{amsfonts}       
\usepackage{nicefrac}       
\usepackage{microtype}      
\usepackage{lipsum}
\usepackage{fancyhdr}       
\usepackage{graphicx}       
\graphicspath{{media/}}     
\usepackage{epigraph}
\usepackage{amsmath}   
\usepackage{amssymb}   
\usepackage{natbib}

\usepackage{siunitx}

\renewcommand{\arraystretch}{1.2}  
\makeatletter
\newcommand{\checknextarg}{\@ifnextchar\bgroup{\gobblenextarg}{}}
\newcommand{\gobblenextarg}[1]{\,\mathrm{#1}\@ifnextchar\bgroup{\gobblenextarg}{}}
\makeatother

\newif\ifstartedinmathmode
\newcommand{\msun}{%
  \relax\ifmmode\startedinmathmodetrue\else\startedinmathmodefalse\fi
  {\ifstartedinmathmode\unit{M_{\odot}}\else$\unit{M_{\odot}}$\fi}\xspace%
}

\newif\ifstartedinmathmode
\newcommand{\rsun}{%
  \relax\ifmmode\startedinmathmodetrue\else\startedinmathmodefalse\fi
  {\ifstartedinmathmode\unit{R_{\odot}}\else$\unit{R_{\odot}}$\fi}\xspace%
}

\makeatletter
\renewcommand\@makecaption[2]{%
  \par
  \vskip\abovecaptionskip
  \begingroup
    \footnotesize\rmfamily
    \begingroup
      \samepage
      \flushing
      \let\footnote\@footnotemark@gobble
      \ifnum\pdfstrcmp{\@captype}{table}=0
        \@make@capt@title{\textsc{Table \thetable}}{#2}%
      \else
        \ifnum\pdfstrcmp{\@captype}{figure}=0
          \@make@capt@title{\textsc{Figure \thefigure}}{#2}%
        \else
          \@make@capt@title{#1}{#2}%
        \fi
      \fi\par
    \endgroup
  \endgroup
  \vskip\belowcaptionskip
}
\makeatother

\defcitealias{Kiel+2009:2009MNRAS.395.2326K}{KH09}

\begin{document}

\author{Dieter Horns\,\orcidlink{0000-0003-1945-0119}}
\author{Niklas Knop\,\orcidlink{0009-0009-4215-5442}}
\author{Mohammad Mohammadidoust\,\orcidlink{0009-0006-5258-9226}}

\affiliation{Institut für Experimentalphysik, Universität Hamburg, Luruper Chaussee 149, D-22761, Hamburg, Germany}

\email{dieter.horns@uni-hamburg.de}
\email{niklas.knop@studium.uni-hamburg.de}
\email{mohammad.mohammadidoust@studium.uni-hamburg.de}

\title{Contribution of Interstellar objects to local dark matter density}

\begin{abstract}
	The recent discovery of three interstellar comets in the solar system indicates the presence of so-far unaccounted baryonic matter in the Galaxy as a population of inter-stellar objects (ISO). The contribution of ISOs to the overall mass budget of the Galaxy affects the estimates on mass of the non-baryonic dark matter halo. We are attempting to estimate the mass density of non-baryonic Dark Matter after including a Galactic ISO contribution to the Galactic rotation curve. 
	The object 3I/ATLAS is a surprisingly massive object with estimates of the
	nuclear radius reaching up to few kilo-metres. The observed incidence rate of interstellar objects (ISO) passing through the inner solar system in combination with estimates on the mass density and size provides an estimate of the local mass density if ISOs in the interstellar medium. The resulting estimate carries large uncertainties which are the consequence of the difficulties to constrain or measure the nuclear radius. The large kinematic age of 3I/ATLAS motivates a model where ISO objects are distributed in a thick (0.8~kpc) disk with a large radial scale length of $\approx 7$~kpc estimated from a fit to rotational velocity measurements from GAIA DR3 data. 
	We find that the ISO contribution to the baryonic mass budget could reach a total mass of $5\times 10^{10}~M_\odot$ which leads to a reduction of the local Dark Matter halo density to $0.24$~GeV/cm$^3$. Even though this
	scenario requires an overly optimistic fraction of matter to be released in the form of ISO objects, it is plausible that the local Dark Matter halo density is biased towards large values given our ignorance of 
	non-detectable baryonic matter in the Galaxy. 

\end{abstract}

\maketitle
\section{Introduction}
\label{intro}
The observed rotation curve of the Galaxy, as well as other spiral galaxies, appears to require the presence of a quasi-spherical halo of unaccounted matter to explain the constant rotational velocity at large galactocentric distances. 
The known baryonic inventory includes the stellar disk, the bulge and gas with a local surface density $\Sigma_{\rm baryon}\simeq54\ M_{\odot}\,\mathrm{pc}^{-2}$.
These components dominate the overall mass budget within the solar galactocentric distance.  
At larger distances, an unaccounted (\textit{dark}) mass component with $\rho_{\text{DM}}(R)\propto R^{-2}$ 
for a galactocentric distance $R$ is required to maintain a constant rotational velocity \citep{read2014local}.  
Commonly, a particle-based form of dark  matter, such as Weakly Interacting Massive Particles (WIMPs), is favoured  as they are well-motivated in the framework of the $\Lambda$CDM standard cosmology.
Baryonic matter in the form \textit{Massive Astrophysical Compact Halo Objects} (MACHOs) has been constrained with observations of gravitational (micro-)lensing events towards the Galactic bulge and the Magellanic clouds \cite{Munoz:2016tmg}.
Initially, these observations have found that objects in the mass range of Pluto- to stellar mass objects do not contribute significantly to the mass budget. 
It is noteworthy that continuing observations of microlensing events  have revealed a population of free-floating planets (FFPs) and isolated black holes \cite{Garcia-Bellido:2024yaz, Nitz:2022ltl, Mroz:2024wia}.
\\
The recent discovery of hyperbolic interlopers in the solar system provides an opportunity to estimate the contribution of these long-lived objects to the overall mass budget in the Galaxy.  
To date, three inter-stellar objects (ISOs) have been directly observed, spanning effective sizes from 600 metres, as in the case of 1I/'Oumuamua \cite{williams2017minor}, up to possibly several kilometres with the recently discovered 3I/ATLAS \cite{seligman2025discovery}. 
By assuming that the stellar Age-Velocity Relation (AVR) could be applicable to the ISO population\cite{kinematic_age_ATLAS}, the high velocity of \SI[separate-uncertainty=true]{57.942(49)}{\kilo\metre\per\second} \cite{seligman2025discovery} observed in 3I/ATLAS suggests this old ($3-11$~Gyrs) population may be distributed in an extended, thick disk inside the Galaxy.
\paragraph{}Recent post-perihelion observations have constrained the radius of 3I/ATLAS  to be between 0.16 and 2.8 km \cite{jewitt2025hubblespacetelescopeobservations} while earlier estimates range from 0.44 - 5.6 km \cite{cloete2025upper}. If we assume a typical cometary geometric albedo of $p_V\approx 0.05$ for its solid nucleus and keep the median absolute magnitude of $H_V\approx 12.4$ in mind, this would yield an uncorrected upper limit on the nuclear radius of 3I/ATLAS: $r_{3\mathrm{I}}\approx (10\pm1)\text{km}$. 

\paragraph{} 
In the following, we provide an estimate of the contribution of ISOs to the
baryonic matter density in the solar neighbourhood.
With additional assumptions on the distribution of ISOs in the Galaxy, we determine a best-fitting model to fit the Galactic rotation curve in the outer Galaxy.

\section{Estimates for the mass density of ISOs in the solar neighbourhood}
\label{section2}
The probability of observing $k$ events given an expected number of events
$\lambda$ is given by the Poissonian distribution :
\begin{equation}
    P(k|\lambda) = \frac{\lambda^k e^{-\lambda}}{k!}.
 \end{equation}

With the observed  $k=1$ of massive ISOs like 3I/ATLAS, we can solve for
$P(\lambda|k=1)=\lambda e^{-\lambda}$ and determine the $90~\%$ inner
percentile bounded by $\lambda \in [0.051,4.743]$. 
The detection of an ISO will be limited by a passage sufficiently close $r_\mathrm{det}\approx 4~\mathrm{au}$ to the sun to activate the comet.
The resulting geometric cross section $\pi r_\mathrm{det}^2$ is increased by the gravitational pull of the sun, such that the effective cross sectional area  is given by

\begin{equation}
    \sigma_\mathrm{eff} = \pi r_\mathrm{det}^2 \left[ 1 + \left( \frac{v_\mathrm{esc}}{v_{\infty}} \right)^2 \right],
 \end{equation}
 where $v_{esc}\simeq$\SI{21}{\kilo\metre\per\second} is the escape velocity of
 the sun at distance $r_\mathrm{det}=3.9$ au \cite{seligman2025discovery}, and $v_{\infty}$ = \SI{57.9}{\kilo\metre\per\second}  is the object's hyperbolic excess velocity. 
The resulting estimate of the local mass density $\rho_{\text{ISO}}$ is given
for an estimated $\lambda$ divided by the effective volume swept by the survey over time $T_{survey}(\simeq 5 \text{yrs})$:
 \begin{equation}
	 \rho_{\text{iso}}(r_{3\mathrm{I}}) = \frac{\lambda \cdot \left( \frac{4}{3} \pi r_{3\mathrm{I}}^3 \rho_{\text{bulk}} \right)}{\sigma_\mathrm{eff} \cdot v_{\infty} \cdot T_\mathrm{survey}},
 \end{equation}
 where $r_{3\mathrm{I}}$ is the radius of the nucleus of 3I/ATLAS, which has been estimated
 to be in the range from $200\text{ m}$ to $10\text{ km}$
 \cite{nucleus_size_HST, Eubanks_2025, forbes2025size, cloete2025upper}, and
 the intrinsic mass density is set to $\rho_{\text{bulk}}$
 = \SI{1.0}{\gram\per\cubic\centi\metre} which is within the possible range
 $\rho_{\text{bulk}}$ = $0.62^{+0.47}_{-0.33}$
 \SI{}{\gram\per\cubic\centi\metre} measured for the comet Tempel 1 \cite{hearn2005bulk}.
 The resulting density of 3I/ATLAS-like objects in the vicinity of the solar system is therefore estimated to be
 %

 \begin{equation}
\begin{split}
   \rho_\mathrm{iso} &= 4\times 10^{-5} \frac{M_\odot}{\mathrm{pc}^3}
   \left(\frac{r_{3\mathrm{I}}}{\SI{2}{\km}}\right)^3 \lambda \\
   &\quad \times \frac{\rho}{\SI{1}{\gram\per\cm^3}}
   \left( \frac{T_\mathrm{survey}}{5~\mathrm{yrs}}\right)^{-1}
   \left(\frac{r_\mathrm{det}}{4~\mathrm{au}}\right)^{-2}.
\end{split}
\end{equation}
 
\begin{figure*}[htb]
  \centering
      \includegraphics[width=\linewidth]{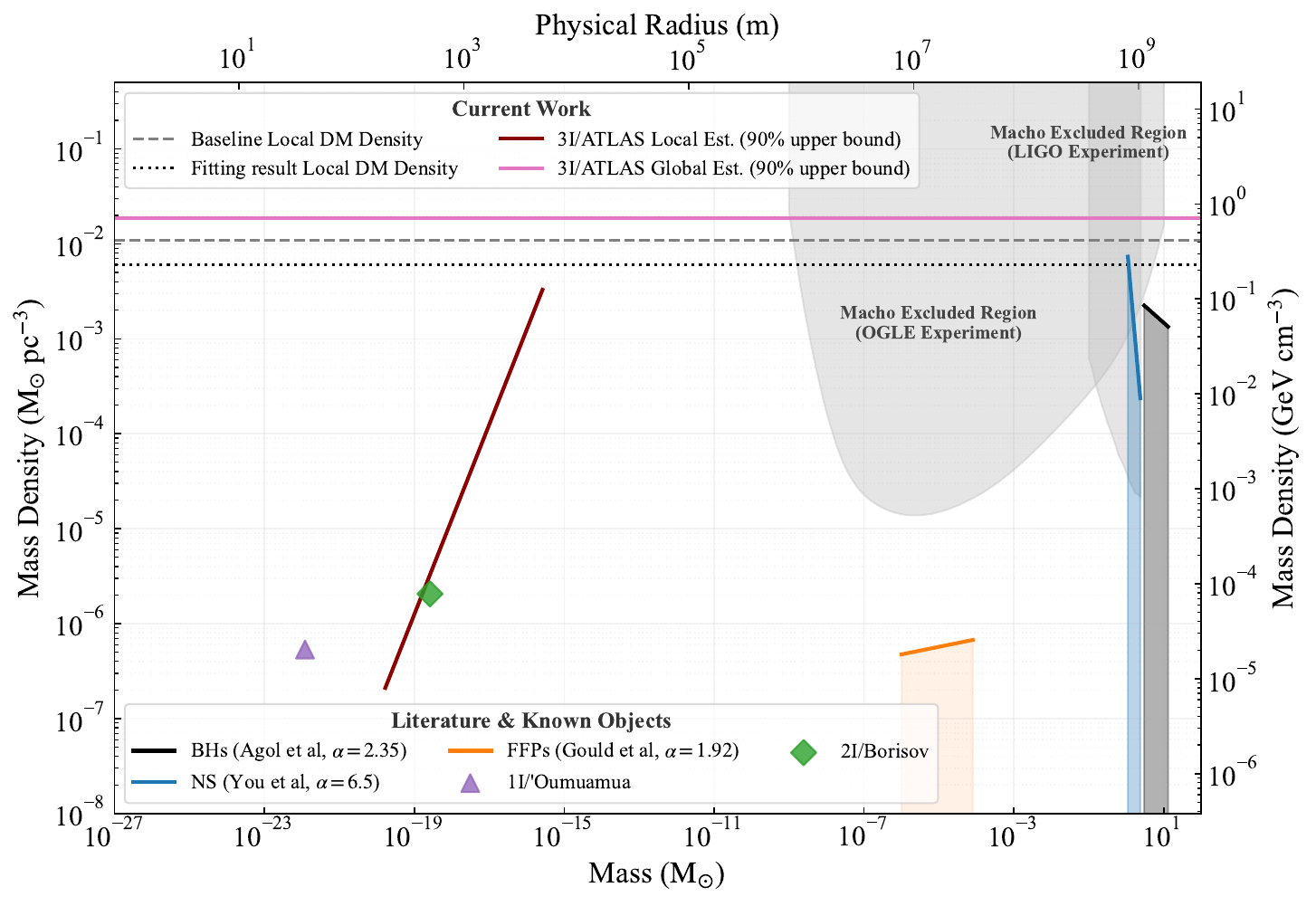}
    \caption{Comparison of 3I/ATLAS mass density with other interstellar
    populations, e.g., isolated black holes \cite{agol2001hb}, neutron stars
  \cite{you2024bmk}, and Neptune-scale free-floating planets
\cite{gould2022free}. Further details are provided in Appendix \ref{appendix:1}. 
The red and pink lines indicate 90\% upper bound on our results based on local and global estimation of 3I/ATLAS-like objects, respectively.
    The dashed and dotted horizontal lines indicate the dark matter values for the
    scenarios introduced in section \ref{section3}. The dashed line represents best-fit value for dark matter density in absence of ISO population, and the dotted line represents the the dark matter density in presence of 90\% upper bound on ISO population fit.}
    \label{fig:mass_density}
\end{figure*}

 The result is shown in Figure \ref{fig:mass_density}, where we compare our local estimate for 3I/ATLAS-like objects with other compact populations and with the densities implied by the Galactic fits. The uncertainty of this local estimate is dominated by the small-number statistics of having detected only one object of this class, which we model with Poisson confidence intervals on $\lambda$. In addition, the inferred density depends sensitively on the assumed nucleus radius, scaling as $\rho_{\mathrm{iso}} \propto r_{3\mathrm{I}}^3$, such that even the modest uncertainties in size translate into large uncertainties in mass density. Further systematic uncertainty arises from the poorly constrained bulk density $\rho_{\mathrm{bulk}}$, the effective detection distance $r_{\mathrm{det}}$, and the adopted survey duration $T_{\mathrm{survey}}$. Therefore, the error shown in Figure~\ref{fig:mass_density} reflects only the statistical uncertainty from the Poisson estimate, while the full physical uncertainty is likely larger. 

\section{ISO distribution in the Milky Way}
\label{section3}
In order to analyse the effect of ISOs contributing to the local dark matter density, it is necessary to extrapolate our local estimations to a global galactic scale.
If a substantial local population of ISOs exists, it must be embedded within a macroscopic density distribution that contributes to the overall rotation curve of the Milky Way.
By constructing a comprehensive mass model and testing it against the Galactic rotation curve, we can determine whether the inclusion of an ISO population remains dynamically consistent with large scale observations and our local estimation.
\subsection{The baseline galactic mass model}
\label{subsection:baseline}

Before we estimate the impact of a Galaxy-wide distribution of ISO objects, we consider a  
baseline model comprising all known baryonic components (gas, stellar populations, and bulge) 
and a standard dark matter halo.
The thin disk and bulge components primarily shape the inner Galaxy's dynamics.
Details regarding the profiles and parameters of these baryonic components are provided in Appendix \ref{appendix:2}.
For the dark matter halo, which dominates the outer Galaxy rotation curve, we have adopted a spherically symmetric Einasto-like 
density profile as suggested by \cite{jiao2023}:
\begin{equation}
    \label{einasto_equation}
    \rho_{\text{DM}}(R) = \rho_{0,DM} \exp \left[ - \left( \frac{R}{h} \right)^{1/n} \right],
\end{equation}

where $\rho_{0,DM}$ is the central density normalisation, $h$ is the radial scale length, and $n$ is the Einasto index. 
Here, we adopted $h = 10~\text{kpc}$, and $n = 1.8$ for the scale length and the Einasto index, respectively.
\paragraph{}Fitting our baseline model to the rotation curve from Gaia DR3 astrometry \cite{jiao2023} provides an estimate of the local dark matter density of $\rho_{DM}(R_0=8.17~\mathrm{kpc})=(0.44\pm0.01)$ GeV/cm$^3$, which is in agreement with recent local and global estimates of $\rho_{DM}(R_0)$ \cite{read2014local}.

For the fitting procedure, the uncertainties on the rotational velocity are limited to the predominantly statistical uncertainty from the split-sample analysis \cite{jiao2023}, ignoring the considerably larger systematic uncertainties.
The resulting best-fitting $\chi^2=8.3$ is therefore comparably small for 17 degrees of freedom.
The best-fitting rotation curve is shown together with the measurement of the rotational velocity for the outer Galaxy in Fig.~\ref{fig:rotation_curve}. 
\subsection{Rotation curve from the ISO population in a thick axisymmetric disk}
The large velocity dispersion of 3I/ATLAS, and therefore its large kinematic age, estimated to be up to $11 \text{ Gyrs}$ indicates that the spatial distribution $\rho_{\text{ISO}}(R, \phi, z)$ of this population could extend well beyond the thin Galactic disk.
As described by \cite{binney_galactic_dynamics}, stochastic gravitational scattering over Giga years increases the random velocities of orbiting bodies, naturally inflating the vertical scale height of the population. 
Consequently, 3I/ATLAS is kinematically inconsistent with the dynamically ``cold'' thin disk (where young stars and gas reside) and is more likely associated with a thick disk.
Following the formalism for  axisymmetric thick disks \cite{binney_galactic_dynamics}, we have assumed that the density distribution $\rho_\mathrm{ISO}(R, z)$ can be factorized into a radial surface density profile $\Sigma(R)$ and a vertical profile $\zeta(z)$:
\begin{equation}
	\rho_\mathrm{ISO}(R,z) = \Sigma(R)\zeta(z).
\end{equation}
We adopt an exponential profile for the radial distribution and a $\text{sech}^2$ profile for the vertical distribution. 
The three-dimensional density distribution is therefore given by:

\begin{equation}
    \rho_{\text{ISO}}(R, z) = \rho_{0} \exp\left(-\frac{R}{R_d}\right) \text{sech}^2\left(\frac{z}{2z_0}\right).
\end{equation}

Here, $\rho_{0}$ is the central volume density normalisation for $R=z=0$. 
The geometric parameters are the radial scale length $R_d$ and the vertical scale height parameter $z_0$. 
In this formulation, the projected surface density $\Sigma(R)$ is related to the volume density $\rho_0$ by integrating the vertical profile over all $z$. 
Using the integral identity $\int_{-\infty}^{\infty} \text{sech}^2(z) \mathrm{d}z = 2$, we obtain the normalisation relation:

\begin{equation}
    \Sigma(R) = 4 z_0 \rho_{0} \exp\left(-\frac{R}{R_d}\right).
\end{equation}

The gravitational potential $\Phi(R, z)$ for such a thick disk is obtained by convolving the potential of a razor-thin disk, $\Phi_0(R, z)$, with the vertical density profile $\zeta(z)$. 
Following the procedure in \cite{binney_galactic_dynamics}, the potential will be obtained by applying a Hankel transformation.
To compute the resulting circular velocity contribution $v_{c, ISO}^2(R)
= R \frac{\partial \Phi}{\partial R}$ in the midplane ($z=0$), the linearity of
the Poisson equation is conveniently used.
The radial force is computed by combining the Hankel transform of the surface density with the vertical form factor. 
For an exponential radial profile, the potential of the thin disk layer $\Phi_0$ is determined by modified Bessel functions ($I$ and $K$). 
The resulting circular velocity squared is:

\begin{equation}
    \label{thick_disc_equation}
    v_{c, ISO}^2(R) = 4\pi G \rho_0 (4z_0) R_d^2 R \int_0^{\infty} dk \frac{k J_1(kR)}{(1 + k^2 R_d^2)^{3/2}} \mathcal{Z}(k),
\end{equation}

where $\mathcal{Z}(k)$ is the Fourier correction factor for the finite thickness of the disk, ensuring the force softens at small scales corresponding to the vertical extent $z_0$. 

\subsection{Fitting Procedure and Results}
In the proposed ISO scenario, the free parameters ($R_d$, $z_0$) of the thick disk are estimated alongside the dark matter halo normalisation from the outer Galactic rotation curve using the same procedure as established for the baseline model (see Sect.~\ref{subsection:baseline}).

\begin{figure}[h]
    \centering
    \includegraphics[width=\linewidth]{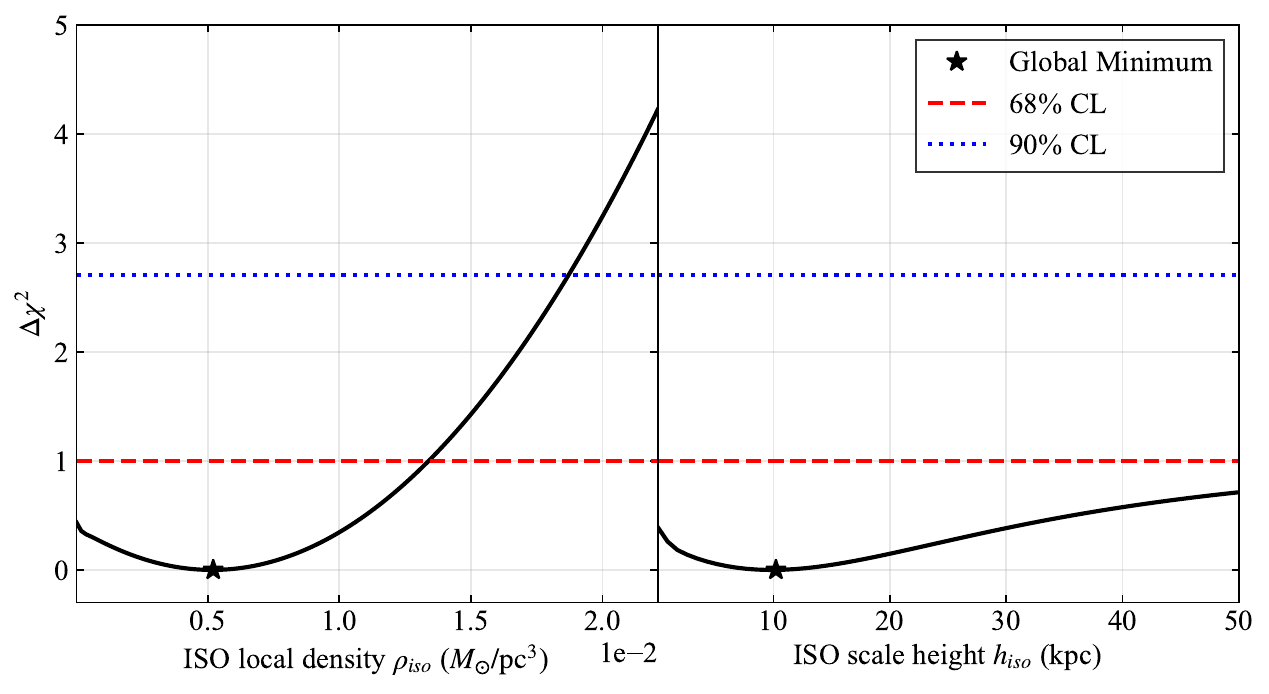}
    \caption{Profile likelihood analysis for the ISO local density ($\rho_{\text{ISO}}$, left) and the vertical scale height ($z_0$, right). The flat profile for $z_0$ demonstrates the insensitivity of the rotation curve to the disk's thickness, necessitating its fixed value in the final fit. The $\rho_{\text{ISO}}$ profile extracts strict $1\sigma$ and $2\sigma$ confidence bounds.}
    \label{fig:profile_fit}
\end{figure}

\paragraph{}While the kinematics of 3I/ATLAS provide an initial physical basis for evaluating ISOs as a thick disk, incorporating this specific model into a rotation curve fit introduces a fundamental parameter degeneracy.
Because the rotation curve is primarily sensitive to the enclosed mass driven by the surface density ($\Sigma \propto \rho_0 z_0$), the volume normalisation $\rho_{\text{ISO}}$ and the vertical scale height $z_0$ become strongly anti-correlated.
As demonstrated in the right panel of Figure \ref{fig:profile_fit}, a profile likelihood scan over the scale height reveals a comparabaly flat test statistic ($\Delta \chi^2 \ll 1$) throughout a large range of values, which indicates that while the radial dependence could be successfully isolated, the rotational data cannot strictly constrain $z_0$ independently of the local density.

To break this degeneracy, we use the vertical Jeans equations to physically bound the scale height with the kinematic properties of 3I/ATLAS.
For a steady-state, axisymmetric galactic disk, the vertical dynamics of a tracer population are governed by the collisionless Boltzmann equation.
Multiplying by the vertical velocity $v_z$ and integrating over velocity space yields the general vertical Jeans equation:

\begin{equation}
  \frac{\partial (\nu \overline{v_R v_z})}{\partial R} + \frac{\partial (\nu \overline{v_z^2})}{\partial z} + \frac{\nu \overline{v_R v_z}}{R} + \nu \frac{\partial \Phi}{\partial z} = 0
\end{equation}

where $\nu$ is the number density of the tracer and $\Phi$ is the total gravitational potential.
By applying some assumptions described in \cite{read2014local, binney_galactic_dynamics} the cross-terms vanish ($\overline{v_R v_z} = 0$).
By converting the number density to mass density ($\rho_{\text{ISO}} \propto \nu$) and identifying the vertical velocity dispersion as $\sigma_z^2 = \overline{v_z^2}$, we obtain a simplified equilibrium condition.
Dividing by $\rho_{\text{ISO}}$, differentiating with respect to $z$, and substituting the vertical component of the Poisson equation for a flattened disk ($\frac{d^2\Phi}{dz^2} \approx 4\pi G \rho_{\text{total}}$), we relate the dispersion directly to the total local mass density:
\begin{equation}
  \sigma_z^2 \frac{d^2 \ln \rho_{\text{ISO}}}{dz^2} = -4\pi G \rho_{\text{total}}
\end{equation}

where $\rho_{\text{total}}$ represents the combined midplane volume density of baryons, dark matter, and the ISO population.
Evaluating the second derivative of our adopted $\text{sech}^2(z/2z_0)$ vertical density profile at the Galactic midplane ($z=0$) yields $-1/(2z_0^2)$.
Substituting this into the constrained Jeans equation provides a direct analytical link between the vertical scale height, the total local midplane density, and the vertical velocity dispersion:
\begin{equation}
  \label{scale_height}
  z_0 = \sqrt{\frac{\sigma_z^2}{8\pi G \rho_{\text{total}}}}
\end{equation}

By assigning the vertical velocity of 3I/ATLAS ($\sigma_z \approx 50$ km/s) as a characteristic kinematic estimator for the entire ISO thick disk, the scale height $z_0$ dynamically locks to the local density at each step of the fitting procedure.
This physical constraint effectively breaks the $\rho_0 - z_0$ degeneracy.
The left panel of Figure \ref{fig:profile_fit} illustrates the resulting profile likelihood for the ISO local density after applying this hydrostatic constraint, successfully establishing a strict $1\sigma$ upper bound on the population.

\begin{figure*}[htb]
    \centering
    \includegraphics[width=\linewidth]{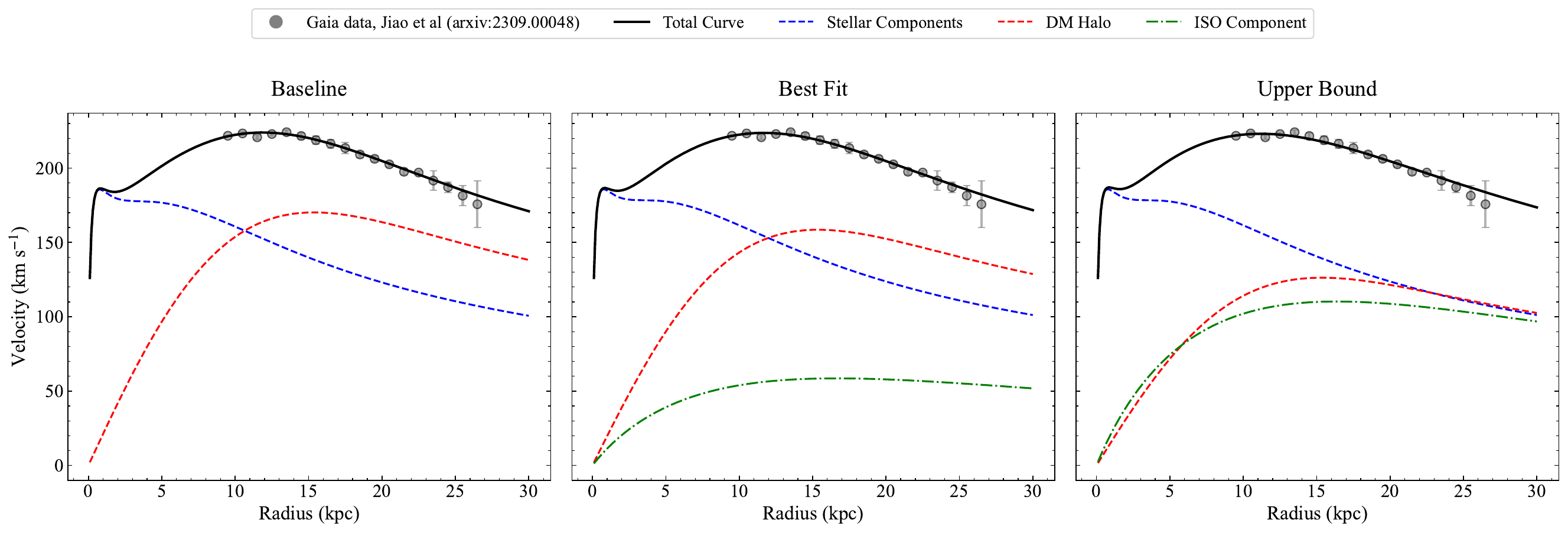}
    \caption{The rotation curve of the Milky Way in three states. The left panel shows the baseline results while the middle and the right panels illustrate the best-fit results and 90\% upperbound on porofiling over local density of ISOs. Here we adopted a total mass of $6.621 \times 10^{10}\,M_\odot$ for all baryonic matters, except ISOs. Observational data is from the Gaia report\cite{jiao2023}}
    \label{fig:rotation_curve}
\end{figure*}

To obtain robust, asymmetric uncertainties for the ISO density, we repeat the profile likelihood analysis for $\rho_{\text{ISO}}$ (Figure \ref{fig:profile_fit}, left panel).
The fitting procedure including the ISO population reduces the overall required normalisation of the dark matter halo by $\approx 13\%$, compensated by an ISO population with a radial scale of $R_d \approx 7.1 \text{ kpc}$. 
The outcome of the fit performed under these two distinct scenarios is summarised in Table \ref{tab:fitresult} and Figure \ref{fig:rotation_curve}.

To validate the goodness-of-fit, we perform a variance analysis on the velocity residuals. We calculate the coefficient of determination ($R^2$), which quantifies the fraction of the raw data's variance successfully explained by the physical model:
\begin{equation}
    R^2 = 1 - \frac{\sum_{i} (v_{obs, i} - v_{model, i})^2}{\sum_{i} (v_{obs, i} - \bar{v}_{obs})^2}
\end{equation}
We find that both the baseline and ISO fit models successfully suppress $\simeq 98\%$ of the raw data's variance, indicating that our parametric models adequately capture the macroscopic Galactic rotation curve.

As depicted, the ISO population is capable of suppressing the local dark matter density by $\sim 13\%$ while maintaining excellent dynamic agreement with the Milky Way's total gravitational potential.

\begin{table*}[htb]
  \centering
  \caption{Summary of fit results for the Galactic mass model. The \textbf{Baseline} scenario assumes a Galaxy composed solely of standard baryonic components and an Einasto dark matter halo. \textbf{ISO Fit} introduces the ISO population as a background potential. To break the physical degeneracy, the vertical scale $z_0$ is dynamically constrained by equation \ref{scale_height}, alongside three other free parameters: the DM normalisation, the ISO normalisation, and the ISO radial scale $R_d$.}
  \renewcommand{\arraystretch}{1.3}
  \begin{tabular}{lccc}
    \hline\hline
    \textbf{Variable} & \textbf{Baseline} & \textbf{ISO Fit} (90\% bound) \\
    \hline
    Local DM Density [GeV cm$^{-3}$] & $0.44\pm0.01$ & $0.38$ ($0.24$) \\
    Local ISO Density [$M_\odot$ pc$^{-3}$] & --- & $\num{5.1e-3}$ ($\num{1.9e-2}$) \\
    ISO Vertical Scale $z_0$ [kpc] & --- & $0.8$ (dynamically constrained by Jeans Eq) \\
    ISO Radial Scale $R_d$ [kpc] &  --- & $7.2 (7.0)$ \\
    ISO Total Mass [kg] & --- & $\num{1.6e10}$ ($\num{5.1e10}$) \\
    $\chi^2$ (d.o.f.) & $8.3\ (\mathrm{d.o.f.}=17)$ & $7.6\ (\mathrm{d.o.f.}=15)$ \\
    \hline
  \end{tabular}
  \label{tab:fitresult}
\end{table*}

\section{Impact on direct and indirect detection experiments}
\label{section4}
The inclusion of a thick disk population of ISOs in the Galactic mass model requires a re-evaluation of the local dark matter density, $\rho_{\text{DM}}$.
As mentioned in Table \ref{tab:fitresult}, accounting for the gravitational potential of the ISO population suppresses the inferred $\rho_{\text{DM}}$ by approximately 13\% (up to 45\%, considering upper bound confidence level) when the ISO parameters are dynamically constrained.
While this correction might appear minor, the precise value of $\rho_{\text{DM}}$ is critical for a wide range of applications in particle physics, astrophysics, and cosmology \cite{read2014local, Bertone:2004pz}.
Particularly, in this section, we investigate this effect on two major dark matter experiments: direct detection searches and the Galactic Centre Excess (GCE) as an indirect search strategy.

In every direct detection experiment, $\rho_{\text{DM}}$ serves as the normalisation factor for the expected flux of dark matter particles passing through Earth-based detectors.
Because the expected recoil rate scales exactly linearly with this local density \cite{Bertone:2004pz, read2014local}, any suppression in $\rho_{\text{DM}}$ implies that current exclusion limits on the WIMP-nucleon cross-section—such as those reported by the LUX-ZEPLIN (LZ) \cite{LZ:2022lsv} and XENONnT \cite{XENON:2025vwd} collaborations would be systematically weakened by the same proportion.
To explicitly quantify this degradation, we performed a model-independent simulation using the \texttt{micrOMEGAs} package \cite{Alguero:2023zol} for an ideal 1-ton fiducial Xenon detector.
By evaluating the spin-independent scattering cross-section for a $30 \text{ GeV}$ dark matter particle, we have found that the ISO population will shift the scattering amplitude up,and therefore weakening the experimental sensitivity limit by approximately 18\%.
This result perfectly supports the expected inverse proportionality driven by the suppression of the local dark matter density, and also confirms that the presence of an ISO disk systematically degrades direct detection constraints.

Furthermore, this density parameter is also essential for indirect searches.
In this case, the expected gamma-ray flux ($d\Phi/dE$) scales quadratically as $\rho_{\text{DM}}^2$ by the "J-factor" quantity, which is defined as the line-of-sight integral of the squared dark matter density ($J \propto \int \rho_{\text{DM}}^2 dl$) \cite{Bertone:2004pz, Abazajian:2014fta}.
The $\rho_{\text{DM}}$ profile is also crucial for constraining the local shape of the Galactic halo \cite{read2014local, Baxter:2021pqo}.
Because the J-factor scales with the square of the density, even a marginal suppression will lead to a compounded systematic uncertainty, that must be considered in the interpretation of indirect dark matter searches.
For a circular region of <1° toward the Galactic Center, our baseline model outcome was an average J-factor of $2.65 \times 10^{22} \text{ GeV}^2 \text{ cm}^{-5}$.
Meanwhile, adding the ISO thick disk suppresses this value to $1.97 \times 10^{22} \text{ GeV}^2 \text{ cm}^{-5}$ in the best-fit case, and to $7.88 \times 10^{21} \text{ GeV}^2 \text{ cm}^{-5}$ considering the upper bound on ISO population.
This reduction highlights the importance of ISO populations. This is especially relevant as modern studies increasingly favour astrophysical origins, such as Millisecond Pulsars (MSPs), over spherically symmetric dark matter templates for the GCE \cite{Manconi:2025ogr}.

\section{Summary}
\label{Disscussion}
The main results can be summarized as follows:
\begin{itemize}
\item[a)] The discovery of 3I/ATLAS indicates the presence of an additional
  baryonic component in the Galaxy.  
\item[b)] The mass density in the solar neighborhood 
  is estimated to be (statistical uncertainties only):
  $\rho_\mathrm{iso}= 4 (+19/-3.9)\times 10^{-5} M_\odot/\mathrm{pc}^3 (r_\mathrm{3I}/2~\mathrm{km})^3$, 
  corresponding to $\rho_\mathrm{iso}= 1.5
  (+7.0/-1.4)~\mathrm{MeV}/\mathrm{cm}^3 (r_\mathrm{3I}/2~\mathrm{km})^3$. Note, the
  cubic scaling with the radius $r_\mathrm{3I}$ of the object. 
\item[c)] Going one step further, we have added a population of ISO objects
  with velocity dispersion of $\sigma\approx 50~\mathrm{km/s}$
  distributed in a thick disk with scale height $z_0=0.8~\mathrm{kpc}
  (\sigma_z/50~\mathrm{km/s})$ and fit the resulting model to the rotation
  curve of the outer Galaxy. 
\item[d)] The resulting fit provides an upper bound on the 
  average density of ISO objects at galactocentric distance $R_0=8.17~\mathrm{kpc}$ of $\rho_{ISO}<(1.9\times
  10^{-2} ) M_\odot\mathrm{pc}^{-3}$ (at 90~\% c.l.), consistent with the local estimate.
\item[e)] The total mass of ISOs present is limited by the fit to the rotation curve to be $<5\times 10^{10}~M_\odot$. 
\item[f)] If the population of ISO objects is close to the maximum value consistent with the rotation curve, the dark matter density would
  be reduced to $0.24~\mathrm{GeV}/cm^3$.
  
\item[g)] The presence of interstellar comets requires a downward correction of non-baryonic dark matter: We estimate a lower bound of non-baryonic dark matter 
  at $0.24~\mathrm{GeV/cm^3}$ with direct implications to the sensitivity of direct and indirect dark matter search experiments and observations.
\end{itemize}

The upcoming all-sky surveys (\cite{dd1rosemary}, \cite{hoover2022population}) will greatly increase the number of known ISO-like objects and will help to establish a more refined model of the spatial distribution
of ISO-like objects in the Galaxy. In combination with the contribution of other compact objects in the Galaxy, the distribution of non-baryonic dark matter will be further constrained.

\appendix
\section{Standard Baryonic components of the Galaxy}
\label{appendix:2}
The local surface density normalisation is fixed at $R_0 = 8.17$ kpc \cite{gravity2019gc}. For the bulge potential, we adopted a Hernquist model with a scale lenght of $a = 0.5$ kpc \cite{hernquist1990bulge}, whereas for the stellar and gas populations, an analytic solution for a razor-thin exponential disk \cite{binney_galactic_dynamics} is used. The necessary parameters are summarized in Table~\ref{tab:baryonic_params}.

The resulting total surface density for baryonic objects is in agreement with the reported value in \cite{read2014local}: $\Sigma_b = 54.2 \pm 4.9$\,M$_\odot$\,pc$^{-2}$.

\begin{table}[htbp]
\centering
\caption{Parameters of the baryonic components adopted for the Galactic mass model.}
\label{tab:baryonic_params}
\renewcommand{\arraystretch}{1.2}
\begin{tabular}{llrl}
\hline
\textbf{Parameter} & \textbf{Symbol} & \textbf{Value} & \textbf{Unit} \\
\hline
Local Galactocentric Distance & $R_0$ & 8.17 & kpc \\
Total Local Baryonic Surface Density & $\Sigma_{bary}(R_0)$ & 54.2 & $M_\odot\,\text{pc}^{-2}$ \\
Stellar Disk Local Surface Density & $\Sigma_{stars}(R_0)$ & 37.2 & $M_\odot\,\text{pc}^{-2}$ \\
Stellar Disk Scale Length & $R_{d,stars}$ & 3.0 & kpc \\
Gas Disk Local Surface Density & $\Sigma_{gas}(R_0)$ & 17.0 & $M_\odot\,\text{pc}^{-2}$ \\
Gas Disk Scale Length & $R_{d,gas}$ & 7.0 & kpc \\
Bulge Mass & $M_{bulge}$ & $9.8 \times 10^{9}$ & $M_\odot$ \\
Bulge Scale Length & $a_{bulge}$ & 0.5 & kpc \\
\hline
\end{tabular}
\end{table}

\section{Local estimates of the mass density related to Black Holes, Neutron Stars, and Free Floating Planets}
\label{appendix:1}
To provide a robust comparison with our local and Galactic estimates of ATLAS-like objects, 
we modeled and highlighted three distinct Galactic populations of compact objects in Figure \ref{fig:mass_density}. 
These populations are generally not considered to be part of  the current baryonic inventory, which amplifies their role in contribution to 
dark matter density of our Galaxy. Instead of relying on isolated detections, which have been significantly increasing in all populations \cite{sahu2025ogle, bogdanov2024magnificent, dong2026free}, we constructed continuous mass functions and converted them into mass densities. 
The quantity plotted for the lines on the y-axis is defined as:

\begin{equation}
	\frac{\mathrm{d}\rho}{\mathrm{d}\log_{10} M} = \ln(10) \cdot M^2 \cdot \frac{\mathrm{d}N}{\mathrm{d}M}
\end{equation}

where $\mathrm{d}N/\mathrm{d}M$ is the differential number density for each population, 
and $\frac{\mathrm{d}N}{\mathrm{d}M} \propto M^{-\alpha}$ with a power-law slope of $\alpha$. 
In order to obtain the local number density for each population, the total number of objects and their spatial Galactic distribution need to be known. 
For BHs and NSs we have followed a thin-disk exponential distribution, where the vertical scale of NSs is considered to be higher than BHs \cite{agol2001hb}.

 Integrating over the exponential thin-disk profile $n(R,z) = n_0 \exp(-R/h_R) \exp(-|z|/h_z)$ yields the relation $N_{tot} = 4\pi h_R^2 h_z n_0$. 
 The local number density then, will be obtained by:

\begin{equation}
 n_{0} = \frac{N_{tot}}{4\pi h_R^2 h_z} \exp\left(-\frac{R_{0}}{h_R}\right)
\end{equation}

where $R_{0}$ represents the local radial distance. For the Neptune-scale FFPs, we adopted a method based on relative stellar density rather than a total Galactic integration, utilising parameters from the microlensing study by \cite{gould2022free}, where the number density is anchored to the local main-sequence stellar density. Detailed values for parameters are provided in Table \ref{tab:parameters}.

\begin{table}[ht]
    \centering
    \caption{Parameters for Local mass density estimation of compact object populations}
    \label{tab:parameters}
    \renewcommand{\arraystretch}{1.2} 
    \begin{tabular}{@{}lllll@{}}
        \toprule
        \textbf{Population} & \textbf{Mass Range} & \textbf{Slope} & \textbf{Normalisation Basis} & \textbf{Scale Height} \\
        & ($M_{\odot}$) & ($\alpha$) & & ($h_z$) \\
        \midrule
        \textbf{Black Holes} & $3.0 - 13.0$\cite{agol2001hb} & $2.35$ \cite{agol2001hb} & $N_{\text{tot}} \approx 1.2 \times 10^8 M_{\odot}$ \cite{agol2001hb} & $300$ pc \cite{agol2001hb} \\
        \textbf{Neutron Stars} & $1.1 - 2.36$ \cite{you2024bmk} & $6.5$\cite{you2024bmk} & $N_{\text{tot}} \approx 10^9 M_{\odot}$ \cite{you2024bmk, agol2001hb} & $1000$ pc \cite{agol2001hb}\\
        \textbf{Free-Floating Planets} & $10^{-6} - 10^{-4}$ \cite{gould2022free} & $1.92$ \cite{gould2022free} & $n \approx 2 \times n_{\text{stars}}$ \cite{gould2022free} & Local Scaling \\
        \bottomrule
    \end{tabular}
\end{table}

\section*{Acknowledgements}
We would like to thank the University of Hamburg clusters of excellence for giving us the change to perform such an in depth analysis of interstellar objects in the context of dark matter.
DH acknowledges support by the Deutsche Forschungsgemeinschaft (DFG, German Research Foundation) under Germany’s Excellence Strategy – EXC 2121 „Quantum Universe“ – 390833306.


\bibliographystyle{aasjournal}
\bibliography{manuscript}{}

\end{document}